\documentclass[twocolumn,showpacs,aps,prl,letterpaper,superscriptaddress,nofootinbib,raggedbottom,nobalancelastpage]{revtex4-2}
\usepackage{graphicx,bm}
\usepackage{amsmath}
\usepackage{xcolor}

\usepackage{color}

\begin{document}
\title{Maximally efficient biphoton generation by single photon decay in nonlinear quantum photonic circuits}
\author{Mikhail Tokman}
\affiliation{Department of Electrical and Electronic Engineering and Schlesinger Knowledge Center for Compact Accelerators
and Radiation Sources, Ariel University, 40700 Ariel, Israel}
\author{Jitendra K. Verma}
\affiliation{Department of Physics and Astronomy, Texas A\&M University, College Station, TX, 77843 USA}
\author{Jacob Bohreer}
\affiliation{Department of Physics and Astronomy, Texas A\&M University, College Station, TX, 77843 USA}
\author{Alexey Belyanin}%
\email{belyanin@tamu.edu}
\affiliation{Department of Physics and Astronomy, Texas A\&M University, College Station, TX, 77843 USA}
\date{\today}
\begin{abstract}
{
We develop a general nonperturbative formalism and propose a specific scheme for maximally efficient generation of biphoton states by parametric decay of single photons. We show that the well-known critical coupling concept of integrated optics can be generalized to the nonlinear coupling of quantized photon modes to describe the nonperturbative optimal regime of a single-photon nonlinearity and establish a fundamental upper limit on the nonlinear generation efficiency of quantum-correlated photons, which approaches unity for low enough absorption losses. 
}
\end{abstract}
\maketitle

{\it{Introduction.---}} 
Future
quantum information systems will inevitably rely on chip-scale integrated photonic circuits for generation and control of quantum states of light \cite{solntsev2017, wang2020, arrazola, sergei, ma2020, galan,smit2019,guo2017, zhao2022}. 
Nonlinear cavities and waveguides supporting spontaneous parametric down-conversion (SPDC) would be particularly attractive for these applications due to their compatibility with mature semiconductor technology. The bottleneck in any such circuit is an extremely low efficiency of nonlinear optical
interactions between single-photon states \cite{guerreiro2014}, which is usually mitigated by using strong classical laser drive fields. While
this is a viable strategy if one needs just a source of single photon or biphoton states, more complicated quantum
information schemes require multiple steps (gates) involving nonlinear couplings between single photons. One way to overcome
the low efficiency challenge is to utilize strong coupling of single photon states with resonant quantum emitters in a
nanocavity \cite{lodahl2015, dovzhenko2018}. Another approach to single-photon nonlinearities which we are going to propose and analyze
here is to utilize critical coupling  \cite{yariv}, or rather its generalization to nonlinearly coupled systems and quantized fields. We
will show that this approach allows one to achieve efficient generation of biphotons by parametric decay
of single drive photons in low-dissipation dielectric photonic circuits, even at the optical nonlinearities typical for standard semiconductor materials. We demonstrate that the {\it nonlinear critical coupling} establishes the fundamental upper limit on the single-photon SPDC, which approaches unity for low enough absorption losses.

For a specific illustration we consider a standard photonic circuit element which
includes evanescent coupling of quantized waveguide modes to the quantized modes of a nonlinear ring cavity as in Fig.~1. The usual treatments based on Heisenberg-Langevin or scattering matrix formalisms \cite{scully, yukani2022, tokman2022} assume a sufficiently weak nonlinearity and some sort of the perturbation expansion to avoid dealing with nonlinear operator-valued equations.  
Here, by extending our recently proposed version of the stochastic Schroedinger equation (SSE) formalism \cite{tokman2020, chen2021, alexey2023} to the quantum field  propagation problems, we were able to consider
strongly nonperturbative nonlinear regimes and include the coupling of all quantized degrees of freedom to their
dissipative reservoirs. Simple analytical expressions for the biphoton
output fluxes are obtained in the practical limit of low reservoir temperature as compared to the optical photon energies.
\begin{figure}[h!]
\centering
\includegraphics[width=6cm,height=5cm]{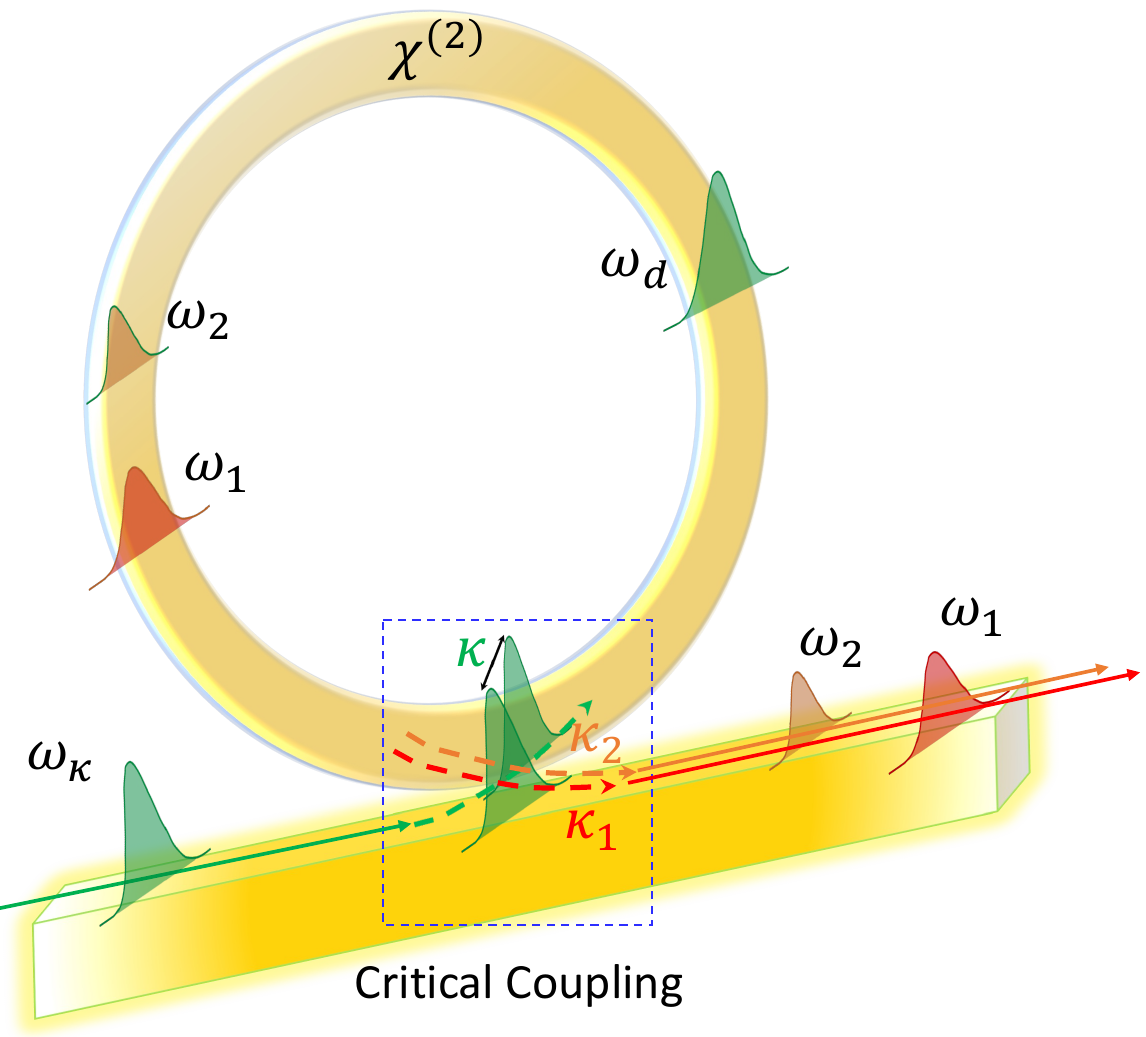}
\vspace{-0.1cm}
\caption{A schematic of a waveguide evanescently coupled to a nonlinear ring cavity made of a crystal with second order nonlinearity $\chi^{(2)}$. A single photon propagates as a wave packet of waveguide modes with frequencies  $\omega_{k}$ and couples to the ring cavity mode at frequency $\omega_{d}$ with coupling coefficient $\kappa$. A drive photon decays parametrically into signal and idler photons $\omega_{1}$ and $\omega_{2}$ which outcouple into the waveguide with outcoupling coefficients $\kappa_{1,2}$.}
\label{fig1}
\end{figure}

Besides the particular application to the nonperturbative SPDC process, our approach develops the critical coupling theory in the fully quantum domain including propagation, dispersion, and dissipation, and fluctuation effects on the same footing, and further generalizes it to include nonlinear optical coupling. One can  apply the same approach to other nonlinear optical processes such as spontaneous four-wave mixing of single-photon modes. Thus our results bring together the burgeoning fields of quantum nonlinear optics, integrated photonics, and telecommunications.

{\it{Quantum Dynamics in the single-photon nonlinear coupling.---}} We begin with expanding all fields into corresponding spatial modes, i.e., (i) waveguide modes $\mathbf{E}_{k}(\mathbf{r}_{\mathbf{\bot}})e^{ikz}$ of  a  waveguide segment with cross-sectional area $\emph{S}$ and length $\emph{L}$,  where $\mathbf{r}_{\mathbf{\bot}}\mathbf{\bot}\mathbf{z}_{0}$, $\mathbf{z}_{0}$ is the unit vector along z axis, and (ii) cavity modes $\mathbf{E}_{N}\left( \mathbf{r}\right)$ of a ring resonator with cross sectional area $S_{R}$ and radius $R$. 
The coupling between waveguide and cavity modes are maximized near the resonance, when $k\approx\frac{N}{R}$ and $\omega_{k}\approx\omega_{N}$ $(\omega_{k,N}$ are certain eigenfrequencies). For boundary-value problems one would need to define the waveguide with open ends and corresponding boundary conditions on the facets. However, since we want to solve the initial value problem, we can start from a certain field wave packet at the initial moment of time inside the waveguide (say, near the boundary) and consider its evolution by expanding it over the
waveguide modes satisfying periodic boundary conditions $k = 2\pi n$, $n = 0, \pm 1, \pm 2\ldots$ This makes sense until the field goes outside the segment, but we can always assume its length $L$ to be large enough. All the results will not depend on the value of $L$.

Proceeding to quantize the fields, one can define the Schroedinger's operator of the drive field in the waveguide as 
\begin{equation}
\hat{\bf{E}}_{wg}({\bf{{r}}}_{\bot},z)=\sum_{k}\big[{\hat{c}}_{k}{\bf{E}}_{k}({\bf{r}}_{\bf{\bot}})e^{ikz} + \hat{c}_{k}^{\dagger}{\mathbf{E}}_{k}^{*}(\mathbf{r}_{\bot}) e^{- ikz} \big],
\label{eqn1}
\end{equation}
and similarly for the cavity field,
\begin{equation}
{\hat{\mathbf{E}}}_{R}\mathbf{=}\sum_{N}\big[{\hat{c}}_{N}\mathbf{E}_{N}(\mathbf{r})\mathbf{+}{\hat{c}}_{N}^{\dagger}\mathbf{E}_{N}^{\mathbf{*}}(\mathbf{r})\big].
\label{eqn2}
\end{equation}
Here {\bf r}  is the position vector in the cavity. The normalization given by Eqs.~({S3}) and ({S4}) in the supplemental material (SM) yields standard bosonic commutation relations for creation and annihilation operators, ${[\hat{c}}_{i}^{\dagger},{\hat{c}}_{j}] = \delta_{ij}$. 

The standard perturbative approach would be to solve for the dynamics of the coupled fields in the Heisenberg's picture starting from the operator-valued Maxwell's equations for the evolution of nonlinearly coupled cavity modes \cite{fain}, 
\begin{equation}
\frac{d\hat{c}_{N}}{dt} + i\omega_{N}{\hat{c}}_{N} = - \frac{i}{\hbar\omega_{N}^{2}}\int_{V}{\mathbf{E}_{N}^{\mathbf{*}}(\mathbf{r}){\ddot{\hat{\mathbf{P}}}}_{ex}(\mathbf{r},t)d^{3}r}
\label{eqn3}
\end{equation}
where ${\hat{\mathbf{P}}}_{ex}$ is an operator of the external polarization which include the nonlinear response and any linear coupling from the outside. Consider three cavity modes, $N = d,1,2$, for which the parametric down-conversion conditions are satisfied:  $\omega_{d} =\omega_{1} + \omega_{2}$ and $N_{d} = N_{1} + N_{2}$. Then the positive-frequency $\propto e^{-i \omega t}$ part of the external polarization is 
\begin{equation}
{\hat{\mathbf{P}}}_{ex;d}^{(\mathbf{+})}= \hat{\chi}^{(2)}(\mathbf{r}, \omega_{d} = \omega_{1} + \omega_{2})\mathbf{E}_{1}(\mathbf{r})\mathbf{E}_{2}(\mathbf{r}){\hat{c}}_{1}{\hat{c}}_{2} + {\hat{\mathbf{P}}}_{lin}^{(\mathbf{+})}, 
\label{eqn5}
\end{equation}
\begin{equation}
\hat{\mathbf{P}}_{ex;1}^{(\mathbf{+})} = \hat{\chi}^{(2)}(\mathbf{r}, \omega_{1} = \omega_{d} - \omega_{2} )\mathbf{E}_{d}(\mathbf{r})\mathbf{E}_{2}^{\mathbf{*}}(\mathbf{r}){\hat{c}}_{d}{\hat{c}}_{2}^{\dagger},
\label{eqn4}
\end{equation}
and similarly for $\hat{\mathbf{P}}_{ex;2}$, where $\hat{\chi}^{(2)}$ is a third-rank tensor of the second-order susceptibility, with permutation properties defined in the SM.  
The operator
${\hat{\mathbf{P}}}_{lin}^{(\mathbf{+})}$  describes linear excitation of the cavity mode at the drive frequency by the waveguide field, which can be parameterized as
\begin{equation}
\int_{V}{\mathbf{E}_{d}^{\mathbf{*}}{\hat{\mathbf{P}}}_{lin}^{(\mathbf{+})}d^{3}r} = \sum_{k}{I_{k}\hat{c}}_{k}.
\label{eqn6}
\end{equation}
Finding the excitation coefficients $I_k$ is a boundary-value problem of the classical electrodynamics. They can be calculated for a specific geometry by any EM solver or conveniently parameterized for any geometry as shown in Sec.~III of the SM, where the conditions for a single-mode excitation are also derived. We can assume that $I_{k}$ reaches a maximum at the drive frequency
 $\omega_{d} = \omega_{k}$ and decays with increasing detuning $|\omega_{d} - \omega_{k}|$. 
 
 To quantify the SPDC process, we introduce the nonlinear overlap integral of the cavity modes with the second-order nonlinearity distribution, integrated over the cavity volume: 
\begin{equation}
G = \int_{V}{\mathbf{E}_{1}^{\mathbf{*}}(\mathbf{r})\hat{\chi}^{(2)}(\mathbf{r}, \omega_{1} = \omega_{d} - \omega_{2})\mathbf{E}_{d}(\mathbf{r})\mathbf{E}_{2}^{\mathbf{*}}(\mathbf{r})d^{3}r}.
\label{eqn7}
\end{equation}
For numerical estimates we will assume that the ring cavity is made of a nonlinear material with a constant value of $\chi^{(2)}$ for given field polarizations. The equations of motion following from Eq.~(\ref{eqn3})  are identical to the Heisenberg equations of motion with the Hamiltonian
\begin{eqnarray}
\hat{H}=&\hbar\sum_{i=d,1,2}\omega_{i}({\hat{c}}_{i}^{\dagger}{\hat{c}}_{i}+\frac{1}{2}) + \hbar\sum_{k}\omega_{k}({\hat{c}}_{k}^{\dagger}{\hat{c}}_{k} + \frac{1}{2})\nonumber\\& - (G{\hat{c}}_{d}{\hat{c}}_{2}^{\dagger}{\hat{c}}_{1}^{\dagger} + h.c.)- \sum_{k}({I_{k}\hat{c}}_{k}{\hat{c}}_{d}^{\dagger}+ h.c.),
\label{eqn8}
\end{eqnarray}
see Eqs.~(S8)-(S11) in the SM for their explicit form.

The next step in a standard approach would be to couple all fields to their dissipative reservoirs and solve the resulting Heisenberg-Langevin equations. However, this approach would lead to operator-valued nonlinear equations \cite{scully, tokman2022} and generally work only in the perturbative regime of low SPDC probability. Instead, we developed the modified version of the SSE formalism which allows us to find analytic or semi-analytic solutions for the nonperturbative quantum dynamics with an arbitrarily strong nonlinearity, determine the maximum nonlinear efficiency, and predict the optimal parameters. 

It is well known from the quantum jumps theory \cite{scully,meystre,Charmichael,zoller,klaus1993,dum,plenio,wiseman} that if the master equation for the dynamics of an open quantum system is  represented in the Lindblad form, $\frac{d\hat{\rho}}{dt} = -\frac{i}{\hbar}[\hat{H},\hat{\rho}] + \hat{\mathcal{L}}(\hat{\rho})$, the Lindbladian can be always written as 
$\hat{\mathcal{L}}(\hat{\rho}) = - \frac{i}{\hbar}({\hat{H}}^{(ah)}\hat{\rho} - \hat{\rho}{\hat{H}}^{(ah)\dagger}) + \delta\hat{\mathcal{L}}(\hat{\rho})$
where ${\hat{H}}^{(ah)}$ is an anti-Hermitian operator.  Then, as we show in \cite{tokman2020, chen2021, alexey2023},  the evolution of the system can be equivalently described by solving the SSE 
\begin{equation}
\frac{d}{dt}|\Psi\rangle = - \frac{i}{\hbar}(\hat{H} + {\hat{H}}^{(ah)})|\Psi\rangle - \frac{i}{\hbar}|\mathfrak{R}(t)\rangle
\label{eqn9}
\end{equation}
where the noise vector satisfies $\overline{|\mathfrak{R}(t)\rangle\langle\mathfrak{R}(t^{\prime})|}=\hbar^{2} \delta {\hat{\mathcal{L}}(\hat{\rho})}_{\hat{\rho}\Rightarrow\overline{|\Psi\rangle\langle\Psi|}} \delta(t-t')$
and the bar means averaging over the reservoir statistics. This approach provides significant analytical and numerical advantages by reducing the effective number of degrees of freedom, especially at the single-photon excitation level. For bosonic modes we have
\begin{equation}
{\hat{H}}^{(ah)} = \sum_{j} - i\hbar\frac{\mu_{j}}{2}\lbrack{\bar{n}}_{\omega_{j}}^{T}{\hat{c}}_{j}{\hat{c}}_{j}^{\dagger} +({\bar{n}}_{\omega_{j}}^{T} + 1){\hat{c}}_{j}^{\dagger}{\hat{c}}_{j}\rbrack, 
\label{eqn10}
\end{equation}
\begin{equation}
\delta\hat{\mathcal{L}}(\hat{\rho}) = \sum_{j = d,1,2} \left[\mu_{j}{\bar{n}}_{\omega_{j}}^{T}{\hat{c}_{j}}^{\dagger}\hat{\rho}\hat{c}_{j} + (\bar{n}_{\omega_{j}}^{T} + 1)\hat{c}_{j}\hat{\rho}{\hat{c}_{j}}^{\dagger}\right],
\label{eqn11}
\end{equation}
where ${\bar{n}}_{\omega_{j}}^{T} = \left(e^{\frac{\hbar\omega_{j}}{k_B T}} - 1 \right)^{- 1}$. To get analytic results, here we consider the case when the photon frequencies are much higher than the thermal energy $k_{B}T$, so that we take ${\bar{n}}_{\omega_{j}}^{T}\rightarrow 0$ in Eqs.~(\ref{eqn10}) and (\ref{eqn11}) and treat the relaxation constants $\frac{\mu_{1,2}}{2}$ as the total field decay rates of the signal and idler cavity modes which include the rate of absorption in the cavity and the rate of cavity field outcoupling into the waveguide: $\frac{\mu_{1,2}}{2} =\frac{\gamma_{1,2}}{2} + \kappa_{1,2}$. The outcoupling rate of the drive cavity field back into the waveguide is already included through the last term in the Hamiltonian (\ref{eqn8}). There is reciprocity between in- and outcoupling rates which we both denote as $\kappa$.

For a single-photon drive, 
the SSE can be solved by the following state vector ansatz:
\begin{eqnarray}
&& |\Psi\rangle = \sum_{k}C_{k}|1_{k}\rangle\prod_{k^{\prime}\neq k}|0_{k^{\prime}}\rangle|0_{d}\rangle|0_{1}\rangle|0_{2}\rangle + \nonumber\\
&& C_{d}\prod_{k}{|0_{k}}\rangle |1_{d}\rangle|0_{1}\rangle| 0_{2}\rangle + C_{12}\prod_{k}|0_{k}\rangle|0_{d}\rangle|1_{1}\rangle|1_{2}\rangle +\nonumber\\  &&  C_{0}\prod_{k}|0_{k}\rangle|0_{d}\rangle|0_{1}\rangle|0_{2}\rangle,
\label{eqn12}
\end{eqnarray}
which results in a set of linear equations for the probability amplitudes, 
\begin{eqnarray}
\frac{dC_{k}}{dt} + i\omega_{k}C_{k} =  \frac{i}{\hbar}I_{k}^{*}C_{d}\label{eqn13}\\
\frac{dC_{d}}{dt} + (i\omega_{d} + \frac{\mu_{d}}{2})C_{d} =  \frac{i}{\hbar}G^{*}C_{12} + \frac{i}{\hbar}\sum_{k}{I_{k}C_{k}}\label{eqn14}\\
\frac{dC_{12}}{dt} + (i\omega_{1} + i\omega_{2} + \frac{\mu_{1}}{2} + \frac{\mu_{2}}{2})C_{12} =  \frac{i}{\hbar}GC_{d}\label{eqn15}.
\end{eqnarray}
Note that the noise vector components are neglected in the equations for the excited state amplitudes  due to the assumption of low reservoir temperature. The noise term must be kept in  the equation for $C_0$, but the latter equation can be omitted  since $C_0$ is not present in other equations and $|C_0|^2$ can be obtained from conservation of noise-averaged norm of the wave vector,  $\overline{\langle\Psi| \Psi \rangle} = 1$ \cite{tokman2020}. Furthermore, we dropped vacuum terms in the free field Hamiltonian, since they do not affect the results which depend only on frequency detunings.  

Consider the initial condition corresponding to a single-photon wavepacket of spectral width $\Delta \omega$ in the waveguide and no photons in the cavity, i.e., $C_{d}(0) = C_{12}(0) = C_{0}(0) = 0$, and $C_{k}(0)\neq0$. The coefficients $I_{k}$ determine the place of the most effective coupling along \emph{z}, which is normally the closest position to the ring cavity. Let all $C_{k}(0)$ are such that at $t = 0$ the field pulse is localized at $z = 0$. 
Although Eqs.~(\ref{eqn13})-(\ref{eqn15}) look like an initial-value problem, in Sec.~IV and V of the SM we show how they include all propagation and linear coupling effects, in particular the critical coupling regime.  


{\it{Signal and idler photon fluxes from single-photon SPDC.---}}  
Eqs.~(\ref{eqn13})-(\ref{eqn15}) are solved for $C_{12}(t)$  with the help of Laplace transforms and the convolution theorem in Sec.~VI of the SM. The result is 
\hspace{-1cm}
\begin{widetext}
\begin{equation}
\left|C_{12}(t)\right|^{2} =\frac{|g|^{2}\left|\Omega_{kd}\right|_{\omega_{k}=\omega_{d}}^{2}}{\left|\kappa_{\Sigma}p_{12} + |g|^{2}\right|^{2}} 
\left\{\sum_{k}{\sum_{q}\left\lbrack C_{k}(0)C_{q}^{*}(0)\left(e^{- i\left(\omega_{k} - \omega_{q}\right)t} + e^{- i\omega_{k}t}U^{*}(t) + e^{i\omega_{q}t}U(t)\right)\right\rbrack} + {\rm c.c.} \right\},
\label{eqn19}
\end{equation}
\end{widetext}
where $$ U(t) = e^{-\left(i\omega_{d} + \frac{\kappa_{\Sigma} + p_{12}}{2}\right)t}
\lbrack\cosh\left(\theta{t}\right) - \frac{{\kappa_{\Sigma} + p_{12}}}{2\theta}\sinh(\theta{t})\rbrack,$$  
$\theta = \sqrt{(\frac{\kappa_{\Sigma} - p_{12}}{2})^{2} - |g|^{2}}$, $p_{12}=\frac{\mu_{1}}{2} + \frac{\mu_{2}}{2} + \omega_{1} + \omega_{2} - \omega_{d}$, and $\kappa_{\Sigma} = \frac{\mu_{d}}{2} + \kappa$ is the total decay rate of the drive field in the cavity which includes its absorption, diffraction, and outcoupling to the waveguide 
modes (see Sec.~V of the SM).  We also introduced the notations $\frac{i}{\hbar}I_{k} = \Omega_{kd}$ and $\frac{i}{\hbar}G = g$. 
Then the photon fluxes outcoupled from the cavity to the waveguide at signal and idler frequencies $\omega_{1,2}$ are (in photons per second)
\begin{equation}
\Pi_{1,2} = 2\kappa_{1,2}\left|C_{12}(t)\right|^{2}.
\label{eqn22}
\end{equation}
The expression (\ref{eqn19}) is still cumbersome, but it can be easily evaluated numerically for a given spectrum $C_k(0)$ of the incident single-photon driving pulse, and its meaning is transparent. The two terms containing $U(t)$ describe decaying or overdamped nonlinear Rabi oscillations between the drive and signal+idler photons, depending on the interplay between the nonlinear coupling strength and the combination of losses and detuning. Note that the terms $e^{- i\left(\omega_{k} - \omega_{q}\right)t}$,
$e^{- i\omega_{k}t}U^{*}(t)$ and $e^{i\omega_{q}t}U(t)$ are of the
same order of magnitude, but the last two terms describing the Rabi
oscillations decay over the timescale no longer than
$\sim\kappa_{\Sigma}^{-1}$, whereas the remaining term 
$\propto \sum_{k}{\sum_{q}{C_{k}(0)C_{q}^{*}(0)e^{-i\left(\omega_{k} - \omega_{q}\right)t}}} +$ c.c.
describes the signal with a much longer duration
$\sim\frac{2\pi}{\mathrm{\Delta}\omega}$, or even an infinitely long
noisy signal with correlation time
$\sim\frac{2\pi}{\mathrm{\Delta}\omega}$. Therefore, for the drive 
pulse which is much longer than the timescales of diffraction and
absorption losses in the ring cavity we can keep only this remaining term. 

The unperturbed incident flux of the drive photons in the waveguide is
\begin{equation}
\Pi_{wg} = \frac{\upsilon_{g}}{2L}\left\lbrack\sum_{k}{\sum_{q}{C_{k}(0)C_{q}^{*}(0)e^{- i\left(\omega_{k} - \omega_{q}\right)t}}} + c.c. \right\rbrack,
\label{eqn23}
\end{equation}
where $\upsilon_{g}$ is the group velocity of the drive wavepacket in the waveguide. Note that in the limit of a large number of modes within the width of the wavenumber spectrum $\Delta k$, when
$\Delta kL \gg 2\pi$, the value of $\Pi_{wg}$ does not depend on the
length of the quantization segment $L$. 
Consider the maximum flux in the middle of SPDC bandwidth at exact frequency matching $\omega_{1} + \omega_{2} - \omega_{d} = 0$. 
Using also 
$\left|\Omega_{kd}\right|^2\simeq  2\kappa\frac{\upsilon_{g}}{L}$ as shown in the SM, the resulting signal and idler fluxes are 
\begin{equation}
\Pi_{1,2} = \frac{8\kappa_{1,2}\kappa|g|^{2} \Pi_{wg}}{\left\lbrack\left(\kappa+\frac{\mu_{d}}{2}\right)\left(\kappa_{1} + \kappa_{2} + \frac{\gamma_{1}}{2} + \frac{\gamma_{2}}{2}\right) + |g|^{2}\right\rbrack^{2}} 
\label{eqn24}
\end{equation}

This is a simple but very informative expression which gives the universal dependence of signal and idler fluxes from all relevant parameters, and includes all possible regimes, such as perturbative (small $|g|$), nonperturbative (arbitrary $|g|$), loss-dominated, and nonlinear critical coupling (see below). Moreover, using more general expressions from the SM it can be generalized to the situations of high reservoir temperatures, broadband and ultrashort pulses, nonlinear Rabi oscillations, etc. It is also important that all the input parameters entering the solution of the quantum nonlinear mixing problem (such as $\kappa$, $g$, $\kappa_{1,2}$, $\mu_d$ and $\gamma_{1,2}$) can be calculated separately, by solving the classical electrodynamics problem with boundary conditions imposed by the photonic circuit geometry and with known material parameters. Therefore, the physics of quantum nonlinear mixing is universal and decoupled from specific experimental geometry.

\begin{figure}[h!]
\centering
\includegraphics[width=8.5cm,height=5cm]{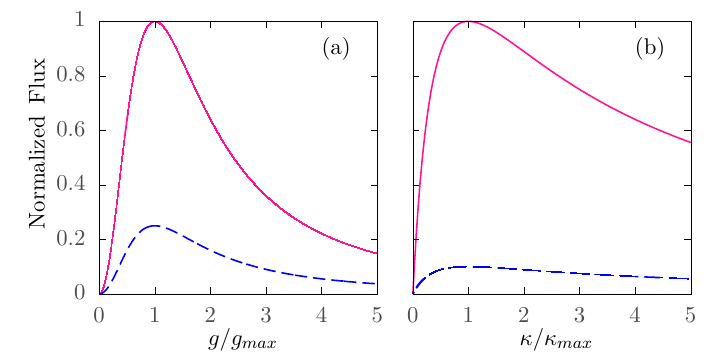}
\vspace{-0.1cm}
\caption{
The normalized flux of signal photons $\frac{\Pi_1}{\Pi_{wg}}$ (a) as a function of the normalized nonlinear coupling strength $g/g_{max}$  for negligible absorption $\gamma_{1,2}=\mu_{d}=0$ (red line) and for $\kappa_{1,2}=\kappa$, $\gamma_{1,2}=\mu_{d}=2\kappa$ (blue-dashed line); and (b) as a function of the normalized linear coupling of the drive field $(\kappa/\kappa_{max})$ for a fixed $g=4.1 \times 10^{8}$ sec$^{-1}$ and  $\kappa=\kappa_{1,2} =g$, without absorption $\gamma_{1,2}=\mu_{d}=0$ (red line), and with absorption $ \gamma_{1,2}=\mu_{d}=2g$ (blue-dashed line).}
\label{fig2}
\end{figure}

\begin{figure}[h!]
\centering
\includegraphics[width=8.5cm,height=5.6cm]{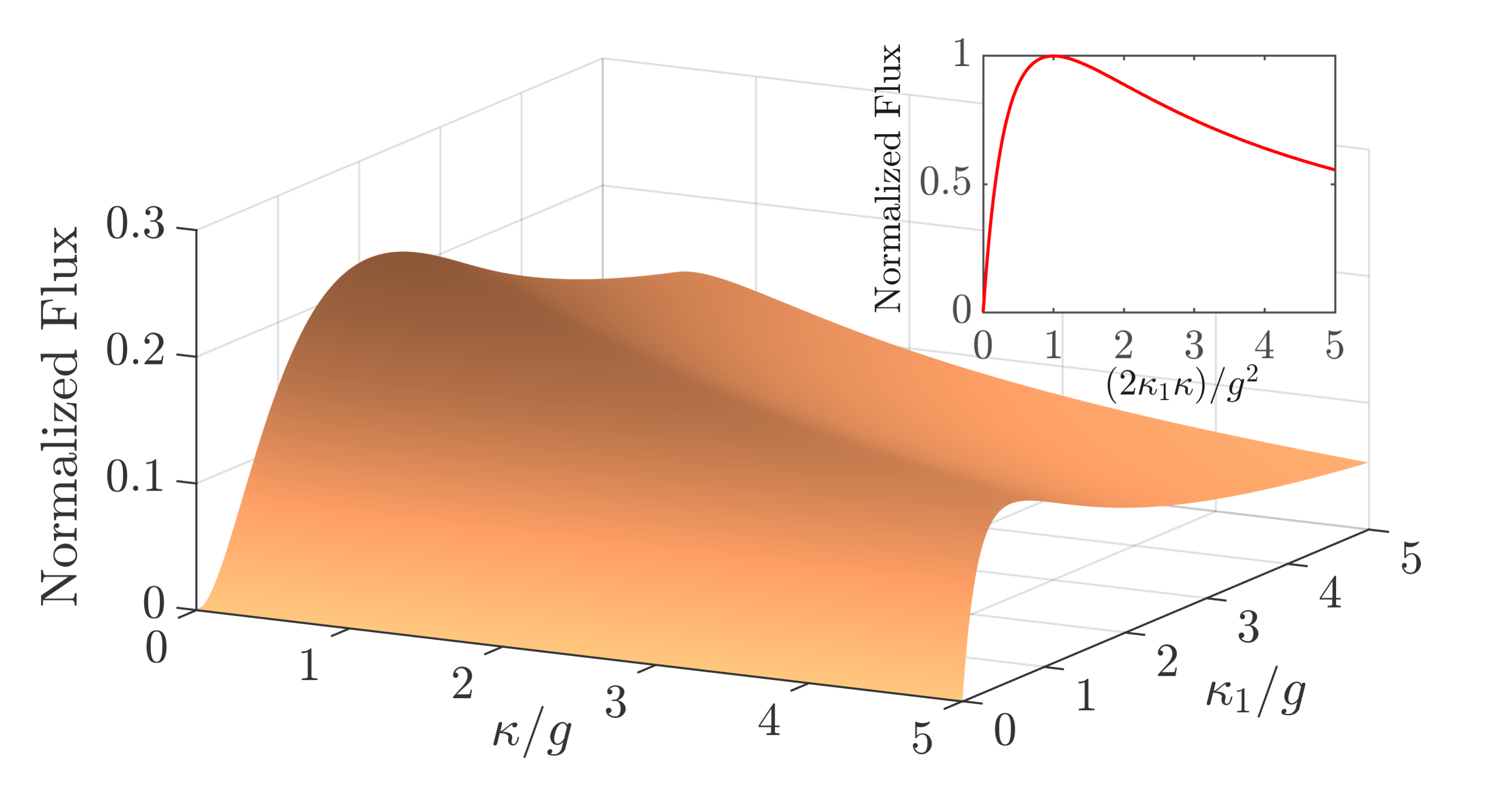}
\vspace{-0.1cm}
\caption{
The normalized flux of signal photons $\frac{\Pi_1}{\Pi_{wg}}$ as a function of the normalized linear coupling strengths of the drive and signal fields for $\gamma_{1,2}=\mu_{d}=g$. Inset: the limit of zero losses, when the flux depends only on the product $\kappa \kappa_{1}$. 
}
\label{fig3}
\end{figure}

Figures 2 and 3 illustrate some of these dependencies for the signal flux normalized by the waveguide flux, $\frac{\Pi_1}{\Pi_{wg}}$, and assuming  degenerate SPDC, $\omega_1 = \omega_2$, $\gamma_1 = \gamma_2$, and $\kappa_1 = \kappa_2$ in order to reduce the number of parameters. The latter assumption can be easily lifted if needed.  The maximum possible signal flux in all figures is of course equal to the incident waveguide flux, i.e., $\frac{\Pi_1}{\Pi_{wg}}=1$, corresponding to one pair of signal and idler photons per one photon at the drive frequency in the waveguide. In agreement with energy conservation, this maximum value is reached when the absorption losses of all fields are much smaller than linear and nonlinear coupling frequency scales. Figure 2(a) illustrates the transition from the perturbative regime $\propto |g|^2$ at small $g$ to the nonperturbative regime of large $g$ as $g = |g|$ changes while keeping other parameters fixed. The maximum flux is reached at  
\begin{equation}
g^{2}_{\max}=\left(\kappa + \frac{\mu_{d}}{2}\right)\left(\kappa_{1} + \kappa_{2} + \frac{\gamma_{1}}{2}+\frac{\gamma_{2}}{2}\right).
\label{eqn25}
\end{equation}
This peak value of the flux is equal to one when absorption losses are negligible (solid red curve), and decreases quickly with increasing losses as is clear from Eq.~(\ref{eqn24})  and illustrated by the blue dashed curve. 

Figure 2(b) shows the dependence of the signal flux from the linear coupling strength $\kappa$ between the waveguide and the cavity at the drive frequency. The peak value of the signal is reached at
\begin{equation}
\kappa_{\max} = \frac{\mu_{d}}{2} + \frac{|g|^{2}}{\kappa_{1} + \kappa_{2} + \frac{\gamma_{1}}{2} + \frac{\gamma_{2}}{2}}. 
\label{eqn26}
\end{equation}
Clearly, this expression is the modification of the {\it critical coupling} condition 
 $ \kappa = \frac{\mu_{d}}{2}$ derived in \cite{yariv} for classical fields and SM for the quantum fields. The significant deviation from standard critical coupling is possible for large enough values of the nonlinear coupling strength $g$. To get a specific numerical estimate for the value of $g$, we consider an InGaP thin-film microring resonator evanescently coupled to a straight waveguide, similar to the type of a nanophotonic circuit implemented in \cite{zhao2022}. Taking the parameters consistent with their devices: a 5 $\mu$m radius and $100\times 400$ nm$^2$ of the ring, $\chi^{(2)} = 220$ pm/V for InGaP \cite{ueno1997}, drive wavelength of 750 nm and signal/idler wavelength of 1500 nm, we obtain $g \simeq 4.1\times 10^8$ s$^{-1}$. For the best possible performance the absorption losses have to be smaller than this value, while the coupling has to be tuned to the critical value (\ref{eqn26}). The absorption losses for the devices reported in \cite{zhao2022} were not quite there yet, with about 1.5\% nonlinearity to loss ratio; however, there is certainly hope that a ratio close to one will be achieved soon. 

Fig.~3 shows the  signal flux when the linear coupling strengths of both the signal and the drive fields are allowed to vary. 
The signal flux reaches its maximum value at the optimal values of the coupling parameters, given by  
\begin{equation}
\kappa_{\max}^2 = \frac{\mu_d}{2\gamma_1} g^2 + \frac{\mu_d^2}{4}; \; \kappa_{1\max}^2 = \frac{\gamma_1}{2\mu_d} g^2 + \frac{\mu_d^2}{4}. 
\label{eqn27}
\end{equation}
This is again the nonlinear critical coupling condition. The maximum value of the normalized flux is given by  
\begin{equation}
\frac{\Pi_{1\max}}{\Pi_{wg}} = \frac{2}{\left[ \sqrt{2 + \mu_d \gamma_1/g^2}  + \sqrt{\mu_d \gamma_1/g^2} \right]^2}.   
\label{eqn28}
\end{equation}
In the limit of negligible losses, the signal flux depends only on the {\it product} of linear coupling parameters, 
\begin{equation}
\frac{\Pi_{1}}{\Pi_{wg}} = \frac{8 \kappa \kappa_1 g^2}{\left( 2 \kappa \kappa_1+ g^2 \right)^2},   
\label{eqn29}
\end{equation}
and its maximum value is of course equal to 1; see the inset to Fig.~3.  

In conclusion, we developed a general nonperturbative theory of second-order nonlinear interactions between quantized field modes in dissipative photonic circuits and proposed a specific realization of maximally efficient SPDC at the single-photon level. Convenient analytic expressions for the biphoton fluxes are obtained which provide explicit dependence on all material and geometric parameters and predict the existence of the nonlinear critical coupling regime which establishes a fundamental upper limit on the signal and idler fluxes for a single-photon pump. 

This work has been supported in part by the Air Force Office for Scientific Research 
Grant No.~FA9550-21-1-0272 and National Science Foundation Award No.~1936276. M.T.~acknowledges the support by the Center for Integration in
Science of the Ministry of Aliya and Integration, Israel.


\end{document}